\documentclass[pdflatex,sn-nature]{sn-jnl}% Math and Physical Sciences Numbered Reference Style 
\usepackage{graphicx}%
\usepackage{multirow}%
\usepackage{amsmath,amssymb,amsfonts}%
\usepackage{amsthm}%
\usepackage{mathrsfs}%
\usepackage[title]{appendix}%
\usepackage{xcolor}%
\usepackage{textcomp}%
\usepackage{manyfoot}%
\usepackage{booktabs}%
\usepackage{algorithm}%
\usepackage{algorithmicx}%
\usepackage{algpseudocode}%
\usepackage{listings}%
\usepackage{caption}
\usepackage{subcaption}
\usepackage{colortbl}

\usepackage{cancel}
\usepackage{enumitem}
\newlist{inlinelist}{itemize*}{1}
\setlist*[inlinelist,1]{label=\textbullet,
                        itemjoin={{ \ }}}
\theoremstyle{thmstyleone}%
\theoremstyle{thmstyletwo}%

\theoremstyle{thmstylethree}%

\begin{document}

\title[Hyperloss]{Hyperloss from coherent spatial-mode mixing in quantum-correlated networks}


\author[1]{\fnm{Stephan} \sur{Grebien}}
\equalcont{These authors contributed equally to this work; the order is determined alphabetically.}

\author[1]{\fnm{Julian} \sur{Gurs}}
\equalcont{These authors contributed equally to this work; the order is determined alphabetically.}

\author[1]{\fnm{Roman} \sur{Schnabel}}

\author*[1]{\fnm{Mikhail} \sur{Korobko}}\email{mikhail.korobko@uni-hamburg.de}

\affil[1]{\orgdiv{Institut für Quantenphysik and Zentrum für Optische Quantentechnologien}, \orgname{Universität Hamburg}, \orgaddress{\street{Luruper Chaussee 149}, \city{22761 Hamburg}, \country{Germany}}}

%\affil*[1]{\orgdiv{Department}, \orgname{Organization}, \orgaddress{\street{Street}, \city{City}, \postcode{100190}, \state{State}, \country{Country}}}


\abstract{
  Quantum-correlated networks distribute quantum resources such as squeezed and entangled states. These states are central to modern quantum technology, including photonic quantum computing, quantum communications, non-destructive biological sensing and gravitational-wave detection. 
Even for squeezed states of light --- the most robust quantum-correlated resource --- loss-induced decoherence remains the dominant obstacle to strong quantum advantage in in large-scale interferometric and networked quantum systems. Common design assumption in these applications is treating mismatches between spatial modes as a small, incoherent loss. Here we show that this picture can fail: coherent spatial-mode mixing with higher-order spatial modes can produce an apparent loss exceeding $100$\,\% relative to the initial squeezing, a regime we term hyperloss.

We experimentally demonstrate hyperloss in a minimal two-node quantum network: with only $8$\,\% mode mismatch, a $5.8$\,dB squeezed state is converted into an effectively thermal state with no quadrature squeezing, eliminating the quantum advantage. Because the effect is coherent, it is controllable: lost correlations can be recovered by tuning differential spatial-mode phases (e.g., Gouy-/propagation-phase). We demonstrate this recovery experimentally, not only eliminating the hyperloss, but even significantly suppressing the mode mismatch loss, with $15$\,\% geometric mismatch acting like only $\approx 2.8$\,\%  effective loss.

Hyperloss is a design-limiting mechanism for all quantum networks with squeezed light, from from photonic quantum processors to large-scale interferometers and distributed quantum-sensing networks.
Our results provide a practical route to avoid hyperloss and turn mode mismatch into an explicit, phase-aware design parameter for future quantum technologies.
}

% \keywords{Quantum optics, Quantum networks, Quantum correlations, Squeezed light, Quantum decoherence, Quantum computing, Quantum communications, Gravitational-wave detection}



\maketitle
% \section{Main}\label{sec:main}

\section{Introduction}
Laser beams with a squeezed quantum uncertainty of the electromagnetic field (``squeezed light'')\,\cite{Walls1983,Breitenbach1997a,Schnabel2017} have become a central resource for quantum technologies and quantum-correlated networks.
They enable breakthroughs in photonic quantum computing~\cite{Fukui2022,Madsen2022,Kashiwazaki2023,aghaeerad2025}, quantum communications\,\cite{Tohermes2025,Conlon2024,Fesquet2024}, and enhance precision interferometry including gravitational-wave detection\,~\cite{Schnabel2010,LSC2011,SchnabelandMcClelland2011,Grote2013,Oelker2014, Tse2019, Acernese2019, ganapathy2023,capoteAdvancedLIGODetector2024,jiaSqueezingQuantumNoise2024, Korobko2025}.
The performance of squeezed states in these quantum-correlated networks is ultimately limited by decoherence. Typically, the dominant degradation channels are optical loss~\cite{Schnabel2017,Korobko2025} and phase noise~\cite{Franzen2006}; corresponding mitigation strategies are well understood, enabling state-of-the-art squeezing levels~\cite{Vahlbruch2016}.
Mode mismatch between spatial modes in an optical network is often approximated as an additional, incoherent loss channel. 
This approximation, however, can drastically underestimate the degradation, as first theoretically recognized in the context of gravitational-wave detection~\cite{Evans2013, Toeyrae2017,Steinlechner2018_Seb, McCuller2021} and more generally applies to multimode interferometric quantum-correlated networks.

We show that in \textit{complex} multimode quantum-correlated networks, coherent mode mixing with strongly anti-squeezed higher-order spatial modes can produce an effective degradation exceeding the full initial squeezing, a regime we call ``hyperloss'', and that it can be mitigated (and in some regimes reversed) by appropriate phase engineering.
Unlike conventional optical loss -- where squeezed light couples to vacuum fluctuations -- hyperloss arises from the spatial mode mixing (SMM) between different transverse modes that are themselves quantum-correlated.
Whenever quantum resources are shared across interfering nodes that support more than one spatial mode, coherent mode mixing can couple the measured mode to correlated (anti-squeezed) degrees of freedom, producing hyperloss.
We experimentally demonstrate that hyperloss can be much more severe than previously appreciated\,\cite{Toeyrae2017, McCuller2021}.
Crucially, we also outline practical mitigation strategies that exploit the coherence of the effect to recover lost correlation, and, in suitable regimes, suppress mismatch-induced degradation by design.

\begin{figure}[t]
    \centering
    \includegraphics[width=1\linewidth]{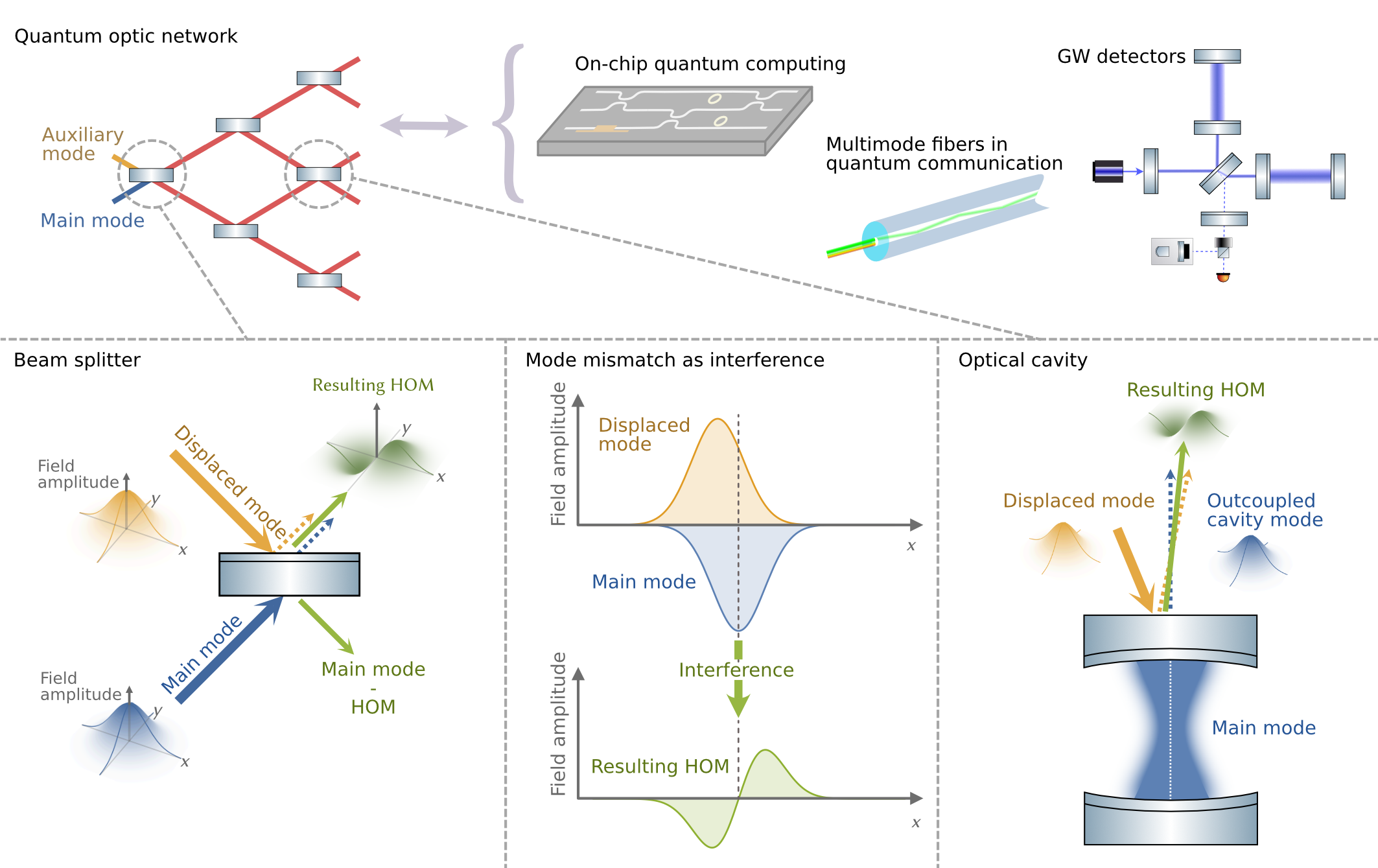}
    \caption{Top: Quantum optical network in a block diagram with multiple nodes that couple several optical modes and three examples of such networks (center to right): a photonic quantum computer, a multimode fiber in a quantum communications or sensing system, and a gravitational-wave detector. Bottom: Two types of coupling nodes that act as spatial mode-mixers: a simple beam-splitter (left) and an optical cavity (right). 
    In both cases, the main mode and the displaced auxiliary mode interfere to produce a higher-order spatial mode upon reflection (center).}
    \label{fig:1}
\end{figure}

Hyperloss sets a scaling limit for quantum-correlated networks that distribute squeezing -- from photonic quantum processors to large-scale interferometers (Fig.~\ref{fig:1}, top). 
So far, it has played a limited role because most existing systems operate with moderate quantum correlations and relatively low network complexity. 
As quantum systems move toward higher squeezing and increasingly complex multimode routing and interference, we expect hyperloss to become a central challenge.
This includes platforms ranging from photonic boson sampling\,\cite{Ng2020} and quantum computing~\cite{Kashiwazaki2023,aghaeerad2025}, to squeezed-light distribution in multimode fibers\,\cite{brieussel2018,leedumrongwatthanakun2020,zuo2025}, and large-scale interferometers~\cite{vermeulen2025,Korobko2025}. 
In this work, we explain the physical origin of hyperloss, demonstrate it experimentally, and outline phase-aware design strategies for future quantum-correlated networks that mitigate this effect.

\begin{figure}[t]
    \centering
    \includegraphics[width=1\linewidth]{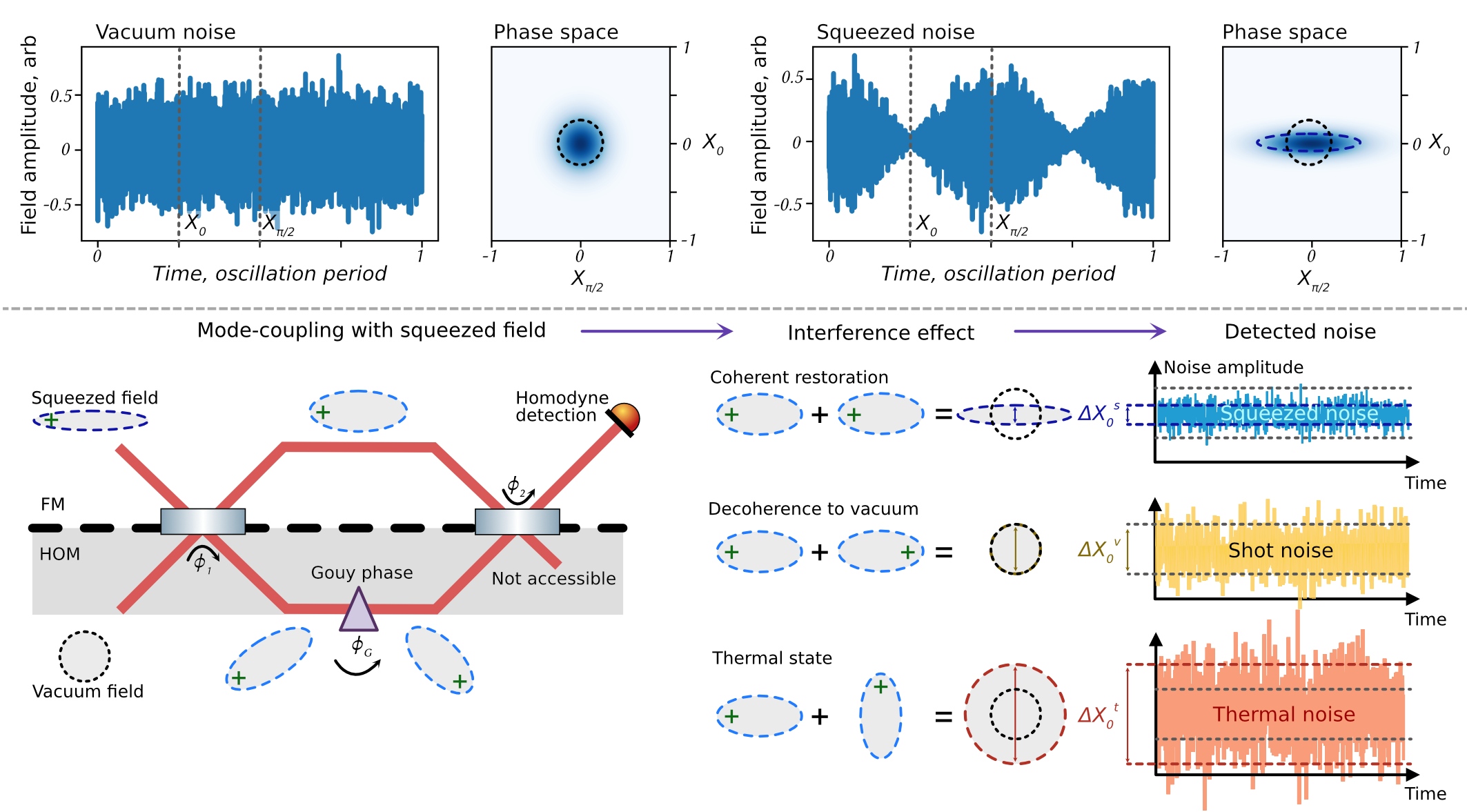}
    \caption{Physical picture behind the hyperloss effect. Top: Field amplitude of the vacuum (left) and the squeezed (right) states as a function of time; and the uncertainty of the measurement record plotted in the phase space with two orthogonal quadratures $\hat{X}_0$ and $\hat{X}_{\pi/2}$.
    Bottom: Mode-coupling leading to interference between different modes and the resulting decoherence effects in the detected noise. 
    The bottom left part demonstrates the mode-mixing at two interference points (cavities or beam-splitters), forming an effective Mach-Zehnder interferometer. Squeezed field in the FM couples to the vacuum field in the HOM. 
    The green crosses are used to keep track of the phase of the state. Upon coupling to the HOM at the first interface, squeezed field acquires a relative phase $\phi_1$, which depends on the parameters of the coupler. After propagation, HOM rotates in phase space due to the Gouy phase $\phi_\text{G}$. Upon coupling at the second interface, FM and HOM have some phase delay, leads to destructive or constructive interference between the coupled fields. Depending on the relative phase, we highlight three special cases (center): when the ellipses end up perfectly aligned, squeezing is fully coherently restored, despite experiencing two mismatches. Fully squeezed noise is observed upon detection. When the ellipses are exactly $\pi$ relative to each other, correlations are completely canceled, and the measured state is at the shot noise level. When the phase is $\pi/2$, anti-squeezing from HOM couples to squeezed quadrature, producing significantly thermal state with increased noise on the detector. This is the hyperloss effect.}
    \label{fig:2}
\end{figure}

\section{Physical origin}

To illustrate the physical origin of hyperloss, we first examine the mechanism of mode mismatch at a beam-splitter and a single-sided (overcoupled) optical cavity, see Fig.~\ref{fig:1}, bottom.
The fundamental mode (FM) of the laser beam is designed to match the auxiliary mode: either on the second port of the beam-splitter, or inside the cavity. 
In practice, however, small imperfections -- either due to deviations in the mode shape, or slight misalignments of the optical axis -- inevitably introduce mode mismatches.
As a result, after the imperfect interference between the two modes, a fraction of the field is scattered into the higher-order spatial modes (HOMs).
Both a beam-splitter and an optical cavity therefore act as couplers between the FM and HOMs.
For small mismatches, this coupling typically involves a single HOM, so they effectively behave as a two-mode beam splitters with one FM and one HOM channels.
This analogy works both for a beam-splitter and an optical cavity, and allows to significantly simplify the treatment of the latter. 
For example, a simple network of two sequentially coupled cavities can be seen to form an effective Mach-Zehnder interferometer, where the FM couples into one HOM and back, see Fig.~\ref{fig:2}, bottom.
Upon the second interaction, two contributions to FM interfere depending on their relative phase, which defines whether the interference is constructive or destructive.
This phase difference arises from two sources: (i) different propagation (Gouy) phase is accumulated by distinct spatial modes~\cite{Feng01}; and (ii) different cavity phase response if two modes are not simultaneously resonant in the cavity\,\cite{Goodwin-Jones2024}.
As a result, the effective loss after propagating through two interfaces becomes significantly different from the baseline loss estimate, which treats SMM as simple, phase-independent optical loss.

We define two regimes of phase-dependent SMM: ``cold'' and ``hot''.
Cold SMM occurs when the light is initially in a coherent state. It is the direct phase-dependent power loss on the FM beam, bounded by 100\,\%.
Hot SMM, on the other hand, appears when the light is in a (pure) quantum-correlated state.
In this regime, the coupling between the spatial modes entangles the fields, and the measurement of one mode yields a mixed, thermal-like state. When interpreted as an effective loss on the squeeze factor, this apparent decoherence exceeds 100\,\%, which motivates the term ``hyperloss''.

In the cold phase-dependent SMM, we can define the effective loss after the two points of mode mixing, see Fig.\,\ref{fig:2}:
\begin{equation}\label{eq:cold}
  \lambda_{\rm smm}\equiv 1- \frac{P_{\rm meas}}{P_{\rm in}} =  2k^2\left(1 + \cos \phi \right),
\end{equation}
where $P_{\rm in, meas}$ are the average measured powers of the input and output light fields, $k$ quantifies the misalignment strength at both cavities (assumed equal), and $\phi$ is the accumulated differential phase between the two modes.
The key feature of the cold phase-dependent SMM can be observed here: mismatch loss $\lambda_{\rm smm}$ cannot be modeled as two independent sources of loss (resulting in $\lambda_{\rm smm} = 1-2k^2$).
Instead, interference between the two coupled modes can either increase or decrease the apparent loss depending on the relative phase $\phi$.
For example, a beam reflected sequentially from two optical cavities, each with $k^2 = 8\,\%$ mode mismatch, would conventionally be assumed to suffer a total loss of about $15\,\%$.
In the cold phase-dependent SMM case the total loss varies between 0\,\% ($\phi = \pi$) and 30\,\% ($\phi = 0$).

In the hot phase-dependent SMM case, the effect becomes even more dramatic, as quantum-correlated light is involved.
Quantum squeezed light is produced in optical nonlinear processes that reduce ("squeeze") the vacuum uncertainty in one quadrature, $\hat{X}_0$, at the expense of increasing the uncertainty in the orthogonal quadrature, $\hat{X}_{\pi/2}$, in accordance with the Heisenberg uncertainty relation, $\Delta^2\hat{X}_0 \Delta^2\hat{X}_{\pi/2}\geq 1$~\cite{Schnabel2022IEEE}, see Fig.~\ref{fig:2} (top).
The level of squeezing is quantified by the squeeze factor $\beta = e^{2r}$, with $r$ the squeeze parameter~\cite{Stoler1971}, defined as the factor by which the variance of the squeezed quadrature is reduced below the ground state level.
Higher squeeze factors, corresponding to higher quantum advantage~\cite{Vahlbruch2016}, are limited by quantum decoherence, most commonly arising from conventional optical loss.
This process can be modeled as a beam-splitter which cross-couples the squeezed state with a ground state, thereby reducing the squeeze factor:
\begin{equation}\label{Eq:loss}
    \Delta^2 \hat{X}_\text{meas}(\Omega) = (1-\lambda_{\rm bs})\Delta^2 \hat{X}_\text{in}(\Omega)  + \lambda_{\rm bs}\Delta^2\hat{X}_\text{vac}(\Omega),
\end{equation}
where $\Delta^2 \hat{X}_\text{in, meas}(\Omega)$ are the uncertainties of the initial and measured decohered state at a Fourier frequency $\Omega$, $\Delta^2\hat{X}_\text{vac}(\Omega)=1$ is the vacuum uncertainty; $\lambda_{\rm bs}$ is the power loss at the beam splitter.
This changes for the SMM process. 
Similarly to the cold SMM, here a part of squeezed field is scattered into the HOM at the first cavity, see Fig.~\ref{fig:2}.
During propagation, the accumulated relative phase causes a rotation of the squeezed quadrature in phase space in one mode with respect to another.
At the second interaction, this phase rotation can mix the anti-squeezed quadrature into the squeezed quadrature of the fundamental field, leading to a severe decoherence effect. 
The total detected noise in the squeezed quadrature can be expressed as:
\begin{equation}\label{Eq: Dh}
     \Delta^2 \hat{X}_\text{meas}(\Omega) = (1-\lambda_{\rm smm})\Delta^2 \hat{X}_\text{in}(\Omega)  + \lambda_{\rm smm}\Delta^2 \hat{X}_\text{vac}(\Omega)  + T(\Omega).
\end{equation}
The term $T(\Omega)$ arises from the projection of the anti-squeezed quadrature from the HOM onto the fundamental mode after the second mixing interface; it scales with the initial anti-squeezing and the mismatch phase $\phi$.
This corresponds to coupling of one quadrature of the field to an effective hot bath.
When we consider the cold SMM ($T=0$) the equation simply reduces to  Eq.~\ref{Eq:loss} with $\lambda_{\rm smm}$ being the source of loss as defined in Eq.~\ref{eq:cold}.
As the thermal contribution increases, the squeeze value reduces even for moderate levels of mismatch, ultimately reaching the point where the state becomes significantly mixed, exceeding the shot noise variance for all quadratures, $\Delta^2 \hat{X}_\text{meas}(\Omega) > 1$. 
The system enters the hyperloss regime.
At the same time, the choice of the total phase between the two modes such that $\Delta \phi = \pi$ allows to recover the lost correlations, making the system effectively immune to the mode mismatch.

\begin{figure}
    %\centering
    \makebox[\textwidth][c]{\includegraphics[width=1\linewidth]{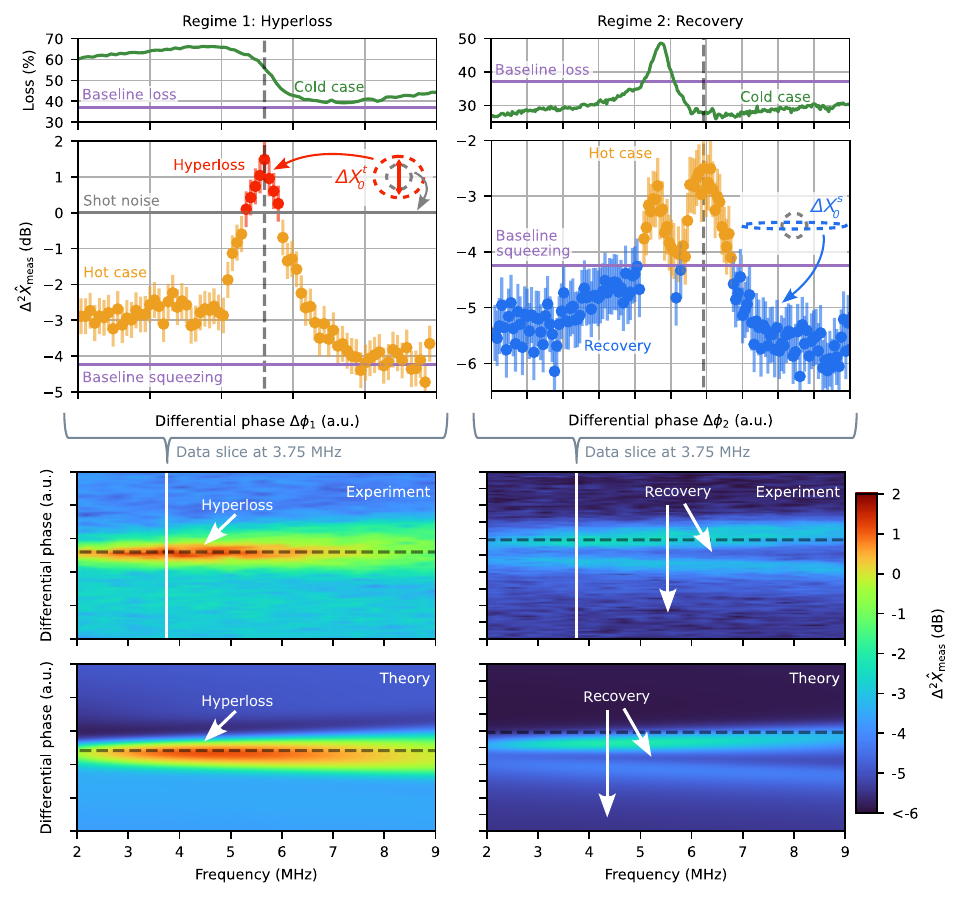}}
    \caption{Top row: power loss from the cold SMM effect on the coherent light field versus the FM-HOM differential phase.
    The baseline (violet) shows the loss level when mode mismatch is treated as direct optical loss.
    The second row: minimal quadrature variance versus the FM-HOM differential phase, relative to shot noise. Two regimes are shown: the hyperloss (left) and recovery (right). In the hyperloss regime, all squeezing is lost and the state is thermal-like with $\approx 1.5\,$dB noise above shot noise level.
    In the recovery regime the noise variance drops below the baseline. 
    Up to 5.2\,dB of initial 5.8\,dB squeezing is recovered, so $\approx 15\,\%$ mismatch acts like only $\approx 2.8\,\%$ effective loss.
    The phase-space illustrations of the measured states are shown as insets, see Fig.\,\ref{fig:2} for details.
    The third row: the full data set, out of which the second row is a slice at 3.75\,MHz (white line). Shown is the minimal noise variance versus measurement frequency and FM-HOM differential phase.
    Bottom row: theoretical simulation captures the observed behavior using independently measured parameters, with the propagation Gouy phase and the detuning of first cavity being the fit parameters.
    }
    \label{fig: result spectrum}
\end{figure}

\section{Experimental demonstration} 

We experimentally demonstrated the transition of the quantum-correlated network system into the hyperloss regime, observing the complete loss of 5.8\,dB of squeezing (\textit{below} the shot noise level) and the transition to an effectively thermal state with $\approx 1.5\,$dB excess noise \textit{above} the shot noise level.
Our setup directly reproduces the physical picture described above, showcasing that even a simple optical network is susceptible to the hyperloss effect.
In our experiment, squeezed field in the $\text{TEM}_{00}$ mode was reflected sequentially from two strongly overcoupled Fabry-Perot cavities, and then measured with a balanced homodyne detector, see Fig.~\ref{fig: detailed setup v2} in Methods for details.
The experienced total optical loss of (26.3 $\pm$ 10.4)\,\%, as measured independently with a well-matched network, allowed to observe up to (5.8 $\pm$ 0.5)\,dB of squeezing and (24 $\pm$ 0.2)\,dB of anti-squeezing at the homodyne detector.
We intentionally introduced an $(8\,\pm1\,)\,\%$ mismatch into the LG$_{01}$ mode at the first cavity while restoring the fundamental mode at the second, so that the HOM existed only between the two cavities.
Since it was not resonant in the either of the two cavities, it acquired a $\pi$ phase with each reflection, which contributed to the overall differential phase between the FM and HOM.

This differential phase was adjusted by two approaches.
i) We chose the phase acquired by the FM upon reflection from the first cavity by detuning it from perfect resonance, with an offset being actively stabilized.
ii) We continuously varied the detuning of the second cavity, imprinting a changing phase shift on the FM upon reflection.
The propagation Gouy phase and other coupling phases were unknown, but this combined dynamical tuning approach allowed us to span the full range of possible phase differences between the two modes.
We performed the full state tomography to record both minimal and maximal noise variance as a function of measurement frequency. We changed the FM-HOM differential phase to characterize different regimes of the hot SMM. We simultaneously recorded the cold SMM effect by propagating a weak coherent beam through the same optical path and measuring the power loss at the homodyne detector.

We highlight two regimes: where the state experiences hyperloss, and where we can cure it by recovering the lost correlations.
Fig.~\ref{fig: result spectrum} presents the minimum of the observed noise variance alongside with theoretical simulations, demonstrating the good qualitative agreement between experiment and theory capturing the complex behavior of the system.
In the hyperloss regime, we observe 1.5\,dB of excess noise above the shot noise level, signifying complete loss of squeezing, see Fig.~\ref{fig: result spectrum}\,(2nd row, left), for 22\,dB of anti-squeezing (see Methods for details).
Remarkably, this effect occurs at a moderate mismatch of only 8\,\%, %\maybe{$8\,\%\pm1\%$},
comparable to the alignment tolerances in many quantum complex optical experiments, highlighting the practical significance of the hyperloss effect.
Importantly, the cold SMM effect, Fig.~\ref{fig: result spectrum}\,(top), cannot explain the observed decoherence, as it shows only 40\,\% of total loss, which would allow to observe up to 4\,dB of squeezing.
This highlights the fundamental difference between the two regimes of phase-dependent SMM and verifies the quantum nature of hyperloss.
We further demonstrate our ability to cure the hyperloss effect by adjusting the differential phase to the recovery regime, see
Fig.~\ref{fig: result spectrum}\,(2nd row, right). We cure up to 12\,\% of loss introduced by the mismatch, recovering the lost correlations. This demonstrates the recipe for curing hyperloss by careful design of the optical network, as discussed below.

\section{Conclusion and outlook}
Our work identifies and experimentally demonstrates hyperloss: a phase-sensitive decoherence regime produced by coherent spatial-mode mixing in quantum-correlated networks.
As quantum optical networks (including photonic circuits, links, and large-scale interferometers) scale in mode count and complexity, hyperloss becomes a practical design constraint unless explicitly mitigated.
For example, current analyses for fault-tolerant CV cluster-state quantum computers (CVQC) conclude that 15-17\,dB of directly measured squeezing in the cluster state is a realistic experimental target for fault tolerance using Gottesman-Kitaev-Preskill (GKP) states and modern error-correction schemes\,\cite{menicucci2014,walsheRobust2019}.
Even with advanced analog error correction that tolerates 5-10\,\% detection loss, fault-tolerant CVQC still requires $\approx$\,10-12\,dB of squeezed GKP resources\,\cite{fukuiHigh2018,Fukui2022,fukui2023}.
Interferometric visibilities for the nodes of such networks in state-of-the-art experiments reach 96-99\,\%\,\cite{asavanant2019}, yet this does not guarantee avoidance of hyperloss when multiple spatial modes can propagate between nodes (for example in free-space sections, cavities, multimode interferometers, or due to residual aberrations).
In a simple extension of our Mach-Zehnder model to a chain of SMM cells, we consider an input squeezing of $15$\,dB and $10$ identical nodes with $1\,\%$ mode mismatch per node. 
The incoherent loss model predicts $10.2$\,dB of squeezing at the output, sufficient for fault-tolerant CVQC.
In the hot SMM limit, where the FM-HOM differential phase is identical in every SMM cell, scanning this phase $\Delta\phi$ over $[0,2\pi)$ shows that for $\approx 55\,\%$ of relative-phase realizations the output falls below a representative $\sim 10$,dB squeeze factor target for GKP-based schemes. 
Although realistic networks will have non-identical phases, this estimation provides an envelope for the effect and highlights the necessity of mitigation strategies.

The coherent nature of hyperloss also enables mitigation.
Hyperloss is highly phase-selective: by tuning differential phases between spatial modes, its impact can be strongly suppressed and, in idealized cases, canceled.
Considering the example above of the chain of SMM cells with a realistic mismatch of $2\,\%$ per node, the incoherent loss model predicts $7.4$\,dB of squeezing at the output, insufficient for fault-tolerant CVQC.
Yet, by optimizing the differential phases between the modes, the output squeeze factor can be maintained at or above the $10\,$dB squeezing target for 25\,\% of the phase values.
This illustrates that phase-aware optical design can not only reduce SMM-induced degradation, but in some regimes exploit coherence to preserve quantum correlations.
Practical routes include: (i) geometric and optical-layout optimization to control Gouy-phase accumulation in different paths; (ii) additional phase-tuning or mode-filtering cavities to compensate unwanted differential phase shifts; (iii) squeezing of HOMs to reduce the impact of hyperloss\,\cite{Steinlechner2018_Seb}; and (iv) active wavefront shaping or adaptive optics to control intermodal phases\,\cite{Goodwin-Jones2024}.

In realistic quantum networks, mismatches rarely involve a single HOM, and each interface would introduce a different set of modes with different coupling strengths.
This makes the design of hyperloss-robust systems more challenging, but the underlying principle remains: by controlling the relative phases between the modes, hyperloss can be mitigated.  
In principle, independently measuring HOM outputs could enable recovery of correlations lost to hyperloss; at scale this would require many additional detectors and substantial optical complexity, making it impractical for many platforms. 
Accordingly, the scalable route is phase engineering through careful Gouy-phase management and cavity detuning design, rather than explicit HOM readout.

We anticipate that hyperloss effects will play a defining role in the design of the next generation of squeezed-light quantum technologies and quantum-correlated networks. Developing proactive, phase-aware multimode design and control strategies to counteract -- and in some regimes exploit -- this effect will be essential for sustaining quantum advantage as these systems scale.

\newpage

\section*{Methods}\label{methods}
Additional experimental details, uncertainty analysis and theoretical derivations are provided in Supplementary Materials.

\subsection*{Experimental setup}

\begin{figure}[H]
    %\centering
    \makebox[\textwidth][c]{\includegraphics[width=1\linewidth]{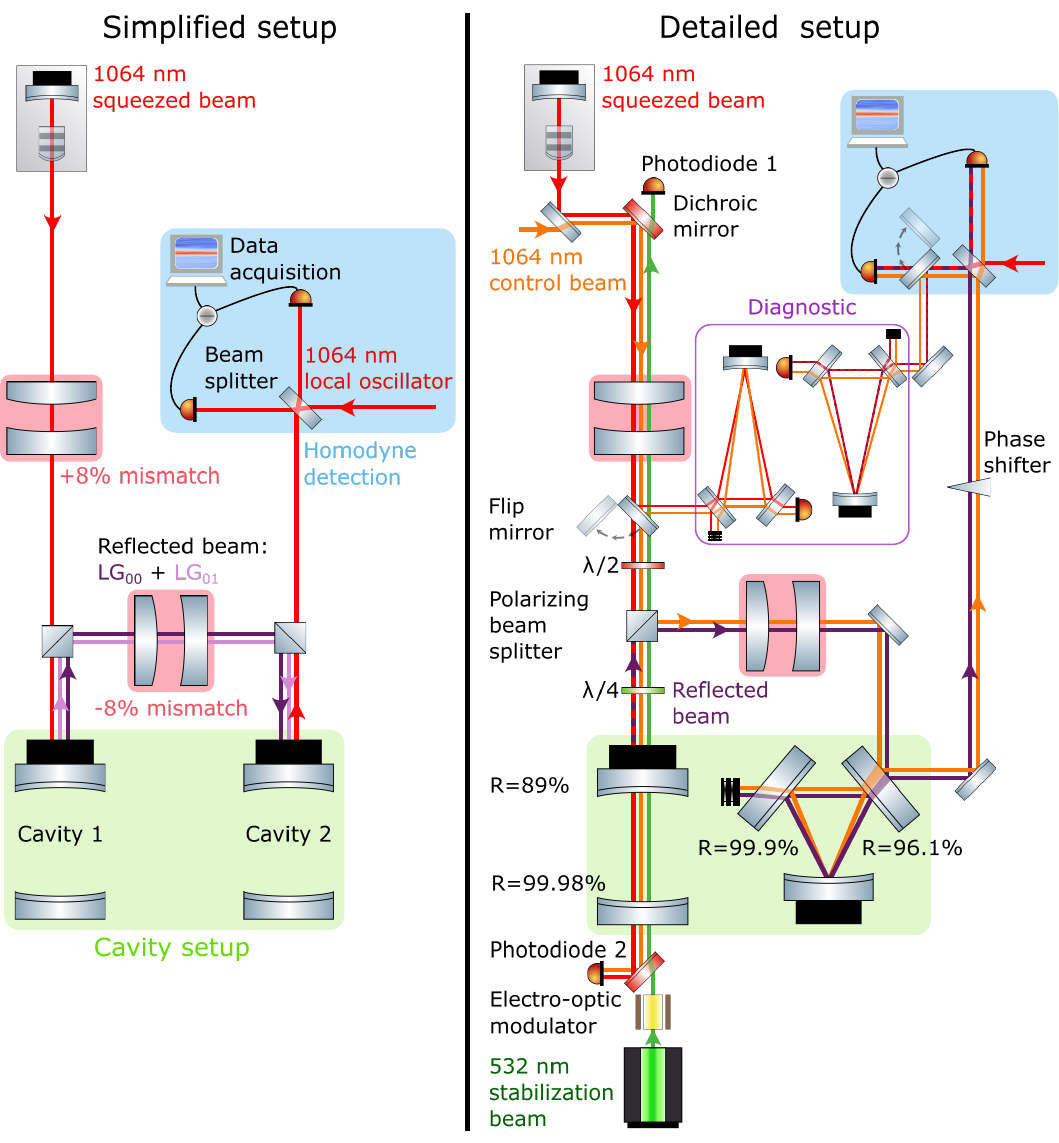}}
    \caption{Left: A simplified setup highlighting the main components of the experiment: the introduced mismatch of $\pm$\,8\,\% (red background), the cavity setup (green background), and the homodyne detection system (blue background). The experiment uses Laguerre-Gauss modes (LG$_{00}$ and LG$_{01}$) to realize the fundamental and higher-order spatial modes. The first set of lenses creates +8\,\% mismatch at a cavity 1, and the second set of lenses compensates it at cavity 2, resulting in pure LG$_{00}$ mode at the homodyne detector.
    Right: A detailed setup, additionally including auxiliary components such as a control beam to assist in aligning the squeezed beam via the diagnostic setup and a stabilization beam for cavity 1.}
    \label{fig: detailed setup v2}
\end{figure}
The experimental setup, shown in Fig.~\ref{fig: detailed setup v2}, consisted of three main parts: the squeezed light source, the two cavities, and the homodyne detection system.
Initially the squeezed beam was prepared in the fundamental spatial Laguerre-Gauss ($\text{LG}_{00}$) mode, with 5.8\,dB of squeezing and 24\,dB of anti-squeezing at 3.75\,MHz, as measured with a well-matched optical network with $\approx 28\,\%$ total loss.
The detailed description of the loss analysis and uncertainties can be found in the Supplementary Materials.
Squeezed field was reflected off two sequentially arranged optical cavities and then measured with a balanced homodyne detector.
Initially the beam was well matched to both cavities with $>99\,\%$ mode-matching efficiency, as verified with independent diagnostic cavities.

Squeezed light was measured with a balanced homodyne detector, where it interfered with a strong local oscillator (LO) beam in the $\text{LG}_{00}$ mode.
The phase of the LO was scanned to measure the full quadrature noise ellipse of the squeezed beam.
The data were recorded using the data acquisition system and then transformed to the frequency domain to obtain the noise spectra as a function of time.

Alongside the squeezed beam, a weak control beam was co-propagated to assist in controlling the squeezed beam, and to serve as classical reference for the cold SMM measurements.
This beam experienced the same mode mismatches and phase shifts as the squeezed beam, allowing us to directly compare hot and cold SMM effects.
We confirmed good overlap between the squeezed beam and the control beam on a separate diagnostic cavity before the main experiment.
The control field was then measured with a DC output of the homodyne detector, observing direct power loss as the change in the maximal observed photocurrent of the interference fringe.

\begin{figure}[H]
    \centering
    \makebox[\textwidth][c]{\includegraphics[width=1\textwidth]{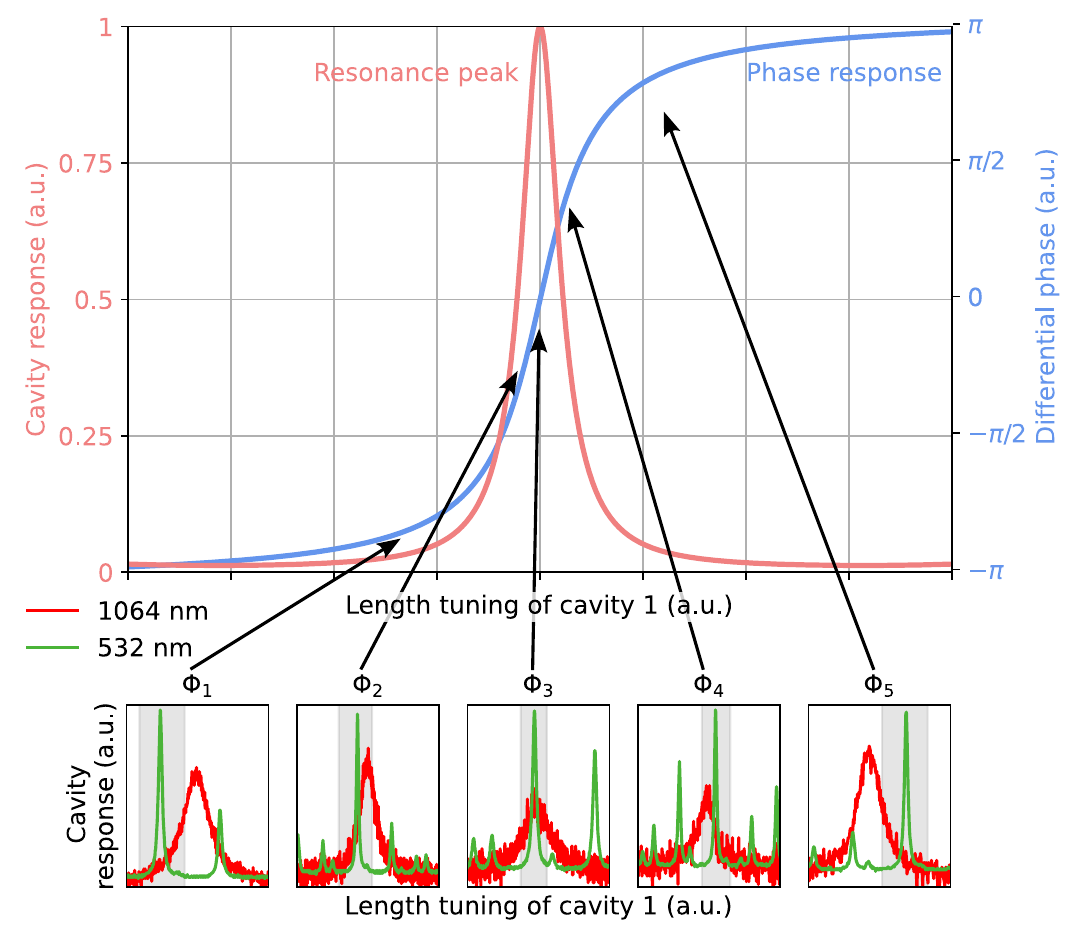}}
    \caption{Top: Phase response of cavity 1's resonance peak at different length stabilization points. Bottom: Resonance peaks of 1064\,nm (monitored at photodiode 2) and 532\,nm (monitored at photodiode 1) for cavity 1. Different modes of 532\,nm were optimized to achieve various differential phases between the fundamental and higher-order modes in response to the 1064\,nm resonance peak.}
    \label{fig: airy peak}
\end{figure}

We introduced a controlled mode mismatch, by using two pairs of lenses before and after the first cavity.
The first pair created a mode mismatch at cavity 1, resulting in coupling of $\approx 8$\,\% of the power into the $\text{LG}_{01}$ mode. Less than 2\,\% of the power was scattered into all other HOMs, neither of which stood out prominently.
We did not include this small contribution in our theoretical model, as it did not significantly affect the results.
The second pair of lenses introduced an opposite mismatch on cavity 2 such, that $\text{LG}_{01}$ mode coupled back in the $\text{LG}_{00}$, resulting in a pure $\text{LG}_{00}$ mode at the homodyne detector, confirmed with the diagnostic cavity.
Thus, the $\text{LG}_{01}$ mode was only present between the two cavities, allowing us to study the SMM effects.

We introduced controlled phase shifts between the fundamental and higher-order modes by two approaches.
First, we detuned cavity 1 from perfect resonance, imprinting a phase shift on the fundamental mode upon reflection, see Fig.~\ref{fig: airy peak}.
This detuning was actively stabilized using a Pound-Drever-Hall locking scheme with a 532\,nm stabilization beam.
Second, we continuously varied the length of cavity 2, changing the phase shift on the fundamental mode upon reflection.
The higher-order mode was not resonant in either cavity, acquiring approximately a $\pi$ phase shift upon each reflection (across the measurement band).
By combining these two tuning methods, we were able to span the full range of possible differential phases between the two modes.

\begin{figure}[H]
\centering
\includegraphics[width=1\linewidth]{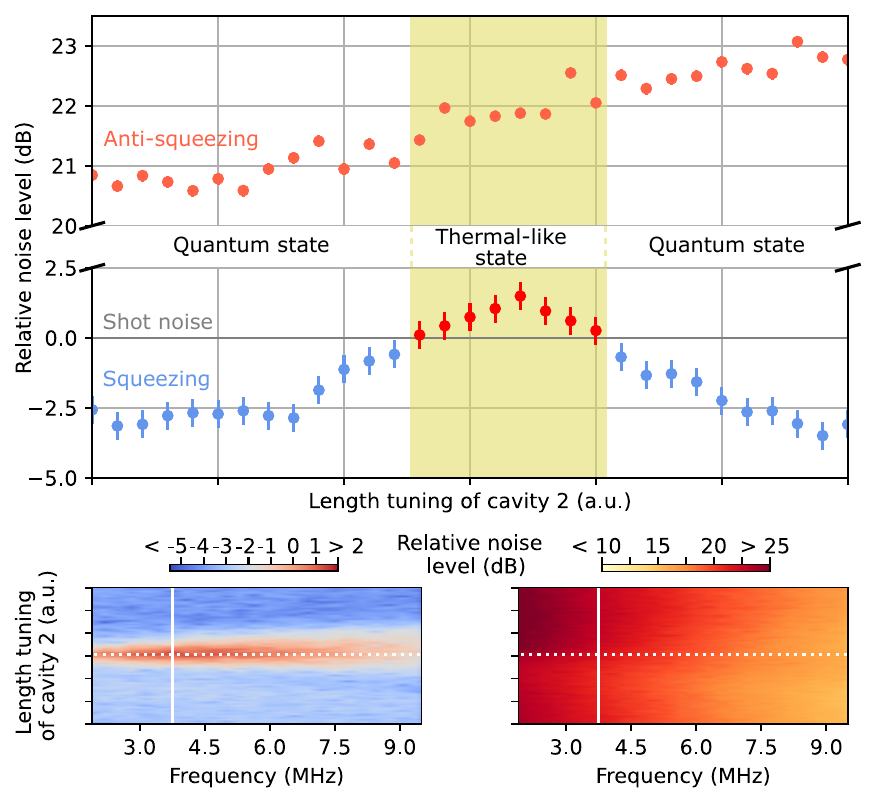}
\caption{Top image: Illustrates the transfer of a quantum state to a thermal state and back by tuning the length of cavity 2 at a measurement frequency of 3.75\,MHz. The quantum states are represented by anti-squeezing (light red) and squeezing (blue). The thermal state is highlighted in yellow. Data points where the squeezing directly evolves into a thermal state are marked in red. The differential phase was set to the hyperloss case, corresponding to the differential phase $\Phi_3$ in Fig.\,\ref{fig: airy peak}. Bottom left: Spectrogram of squeezing. The vertical white line indicates the point of the slice corresponding to the top image, while the dashed horizontal white line marks the position of cavity 2 with the most hyperloss. Bottom right: Spectrogram of anti-squeezing.}
\label{fig: squeezing}
\end{figure}

The hyperloss effect, presented in Fig.~\ref{fig: result spectrum} of the main text, was observed by measuring the variance of the squeezed beam while continuously varying the length of cavity 2 and the homodyne phase at different time scales.
While the squeezed quadrature acquired significant frequency dependence due to the cavity response, we compensated for that by selecting a homodyne angle with a minimal noise variance at each frequency. This ensured that we always measured the quadrature with the least noise, allowing us to accurately capture the hyperloss effect across the entire frequency range.
At the same time, we could monitor the anti-squeezed quadrature by selecting the orthogonal homodyne angle, as shown in Fig.~\ref{fig: squeezing}, bottom.
Depending on the differential phase between the two modes, squeezed quadrature coupled more or less excess noise from the anti-squeezed quadrature, leading to the observed hyperloss effect, where squeezing was completely lost. 
Anti-squeezed, on the other hand, remained mostly unaffected by the phase changes, as shown in Fig.~\ref{fig: squeezing}, top, consistent with the physical picture presented in Fig.~\ref{fig:2} of the main text.

\subsection*{Simplified theoretical model}
To model the SMM effect, we consider a simple Mach-Zehnder-like coupling setup, as shown in Fig.~\ref{fig:2} of the main text.
We assume that the fundamental mode (FM) is initially in a squeezed state, while the higher-order mode (HOM) is in the vacuum state.
The two modes interact at two beam-splitters, representing the mode-mixing interfaces (e.g., optical cavities).
Between the two beam-splitters, the HOM acquires a phase shift $\phi_1$ relative to the FM, which includes both the Gouy phase and any coupling-induced phase shifts.
We denote the coupling strengths at the first and second beam-splitters as $k_1$ and $k_2$, respectively.

With the full derivation provided in the Supplementary Material, we arrive at the following expression for the measured noise spectral density of the measured phase (squeezed) quadrature after the second beam-splitter:
\begin{multline}
  S_{y, {\rm meas}} = \left(\cos k_2 \sin k_1 + \cos k_1 \cos \phi_1 \sin k_2\right)^2 + \\ 
  e^{-2 r_s} \left(\cos k_1 \cos k_2 - \cos \phi_1 \sin k_1 \sin k_2\right)^2 + 
  \cos^2 k_1 \sin^2 k_2 \sin^2 \phi_1 +\\ 
  e^{2 r_s} \sin^2 k_1 \sin^2 k_2 \sin^2 \phi_1.
\end{multline}
We highlight three special cases $\phi_1 = \{0,\pi/2,\pi\}$, corresponding to Fig.~\ref{fig:2}, assuming equal mismatch at two interfaces, $k_1 = k_2$:
\begin{align}  
  & S_{y, {\rm meas}}(\phi_1 = \pi) = e^{-2 r_s},\\
  & S_{y, {\rm meas}}(\phi_1 = 0) = \sin^2 2k_1 + e^{-2r_s} \cos^2 2k_1,\\ 
  & S_{y, {\rm meas}}(\phi_1 = \pi/2) = e^{-2 r_s} \cos^4 k_1 + 2 \cos^2 k_1 \sin^2 k_1 + e^{2 r_s} \sin^4 k_1.
\end{align}
In the first case, squeezing is perfectly coherently restored, despite two mismatches — this is the restoration effect.
In the second case, squeezing is still present, although is significantly suppressed: e.g. for the mismatch of $8\,\%$, $S_{y, {\rm meas}}\approx \sqrt{0.08} (1+2.5 e^{-2r_s})$.
Finally, the third case is most dramatic, where anti-squeezing couples directly into the squeezed quadrature. For example, for the mismatch of $8\,\%$:
\begin{equation}
  S_{y, {\rm meas}}(\phi_1 = \pi/2, k_1 = \sqrt{0.08}) \approx 0.01 e^{2r_s},
\end{equation}
where we assumed that anti-squeezing is strong enough, $e^{2r_s}\gg1$. In this case, we observe complete decoherence of the quantum state for any input squeezing stronger than 20\,dB, for which $e^{2r_s}=100$. Higher input squeezing or stronger mode coupling leads to hyperloss -- a significantly mixed state.

\begin{figure}
    \centering
    \includegraphics[width=1\linewidth]{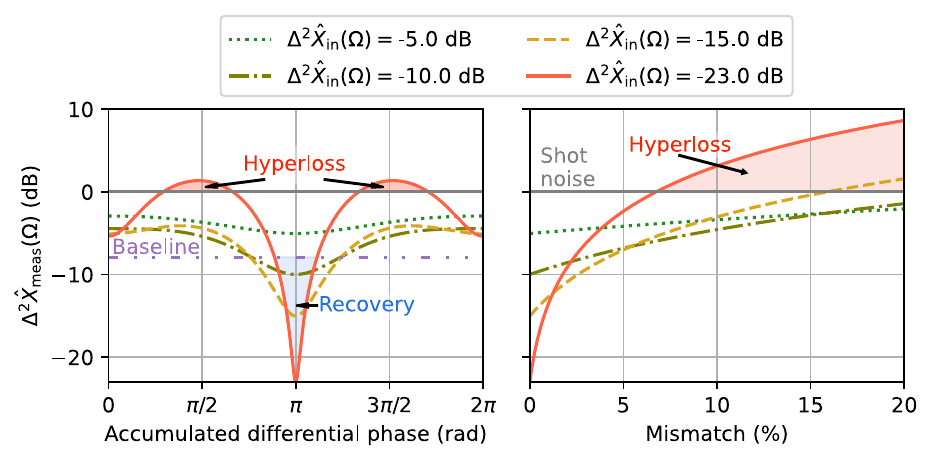}
    \caption{ The effect of hot SMM on a squeezed state. Left: Noise reduction below the shot noise level as a function of the phase difference between the FM and HOM, for different levels of initial squeezing. The effect of hyperloss is visible for $\pi/2$ phase difference, and the effect of recovery -- for $\pi$ phase difference. The mode mismatch is equal to 8\,\% on two nodes. Right: Noise reductio below the shot noise level as a function of the mode mismatch, for different levels of initial squeezing and $\phi=\pi/2$. The higher the initial squeezing is, the less tolerant the system is to the mismatch, entering the hyperloss regime for moderate amounts of mismatch.}\label{fig:theory_full}
\end{figure}

We also consider the case of a weak coupling, where $k_{1,2}\ll1$. We expand sine and cosine up to the second order in $k_1,k_2$, and up to the fourth order for the term proportional to $e^{2r_s}$, since it can be significant if anti-squeezing is strong enough:
\begin{align}
  &S_{y, {\rm meas}} \approx e^{-2r_s}(1-\lambda_{\rm smm}) + \lambda_{\rm smm} + T,\\
  &\lambda_{\rm smm} = k_1^2 + k_2^2 + 2k_1k_2\cos \phi_1,\\
  &T = k_1^2 k_2^2 e^{2r_s} \sin^2 \phi_1,
\end{align}
where we introduced the effective loss $\lambda_{\rm smm}$ and the effective normalized noise temperature $T$. This equation is leads to Eq.\,\ref{Eq: Dh} in the main text.
We again highlight three special cases:
\begin{align}  
  & \phi_1 = \pi:\qquad \lambda_{\rm smm} = (k_1-k_2)^2,\quad T = 0,\\
  & \phi_1 = 0: \qquad \lambda_{\rm smm}=(k_1+k_2)^2,\quad T = 0,\\
  & \phi_1 = \pi/2:\qquad \lambda_{\rm smm} = k_1^2+k_2^2,\quad T=k_1^2 k_2^2 e^{2r_s}.
\end{align}
In the last case, the anti-squeezing is coupling into squeezed field, leading to severe decoherence and creation of a mixed thermal-like state.
Fig.\,\ref{fig:theory_full} shows the dependence of the measured squeezing on the phase difference and the mode mismatch and highlights both hyperloss and recovery effects.

\subsection*{Network analysis}
%as well as biological~\cite{Frascella2021,Li2021} and medical sensing~\cite{Nolte2012Book}, and opens the possibilities for novel dark matter searches~\cite{Backes2021,Shi2023}. It plays an important role in quantum communications~\cite{Tohermes2025,Conlon2024,Fesquet2024}, and in recent experiments demonstrating quantum computational advantage in photonic quantum computers~\cite{Madsen2022,Kashiwazaki2023,aghaeerad2025}.
%Squeezed states have a number of advantages: they are proven to be optimal for sensing applications~\cite{Taylor2016,Nolte2012Book,Krause2012,forstner2012,GW150914}, allow for high experimental rates~\cite{Tohermes2025} and can be created and controlled in a straightforward and well-understood way~\cite{Wu1986,Vahlbruch2010,Vahlbruch2016}.
\begin{figure}
  \centering
  \begin{minipage}{0.5\linewidth}
      \includegraphics[width=1\linewidth]{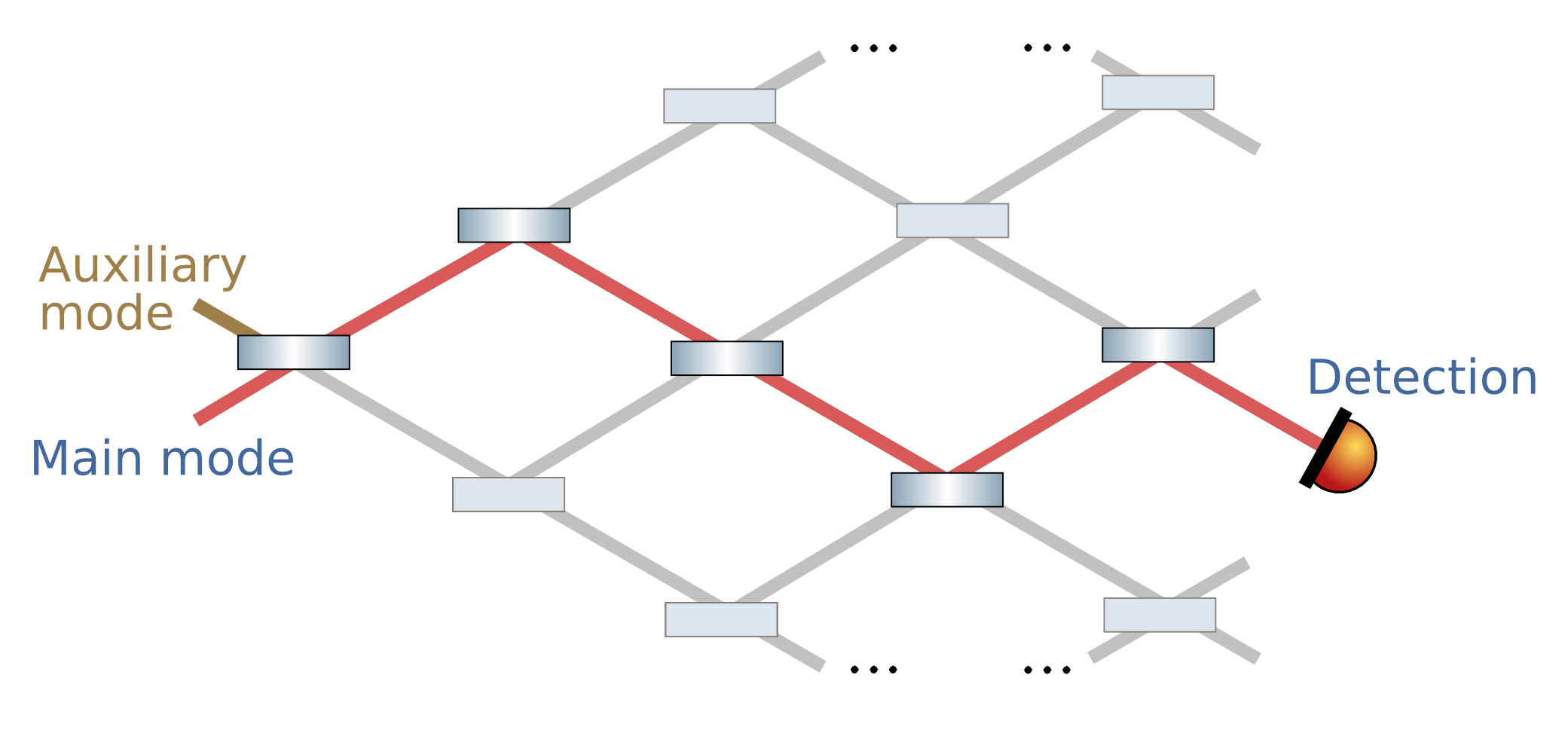}
  \label{fig:theory_network}
  \end{minipage}
  \begin{minipage}{0.45\linewidth}
      \includegraphics[width=1\linewidth]{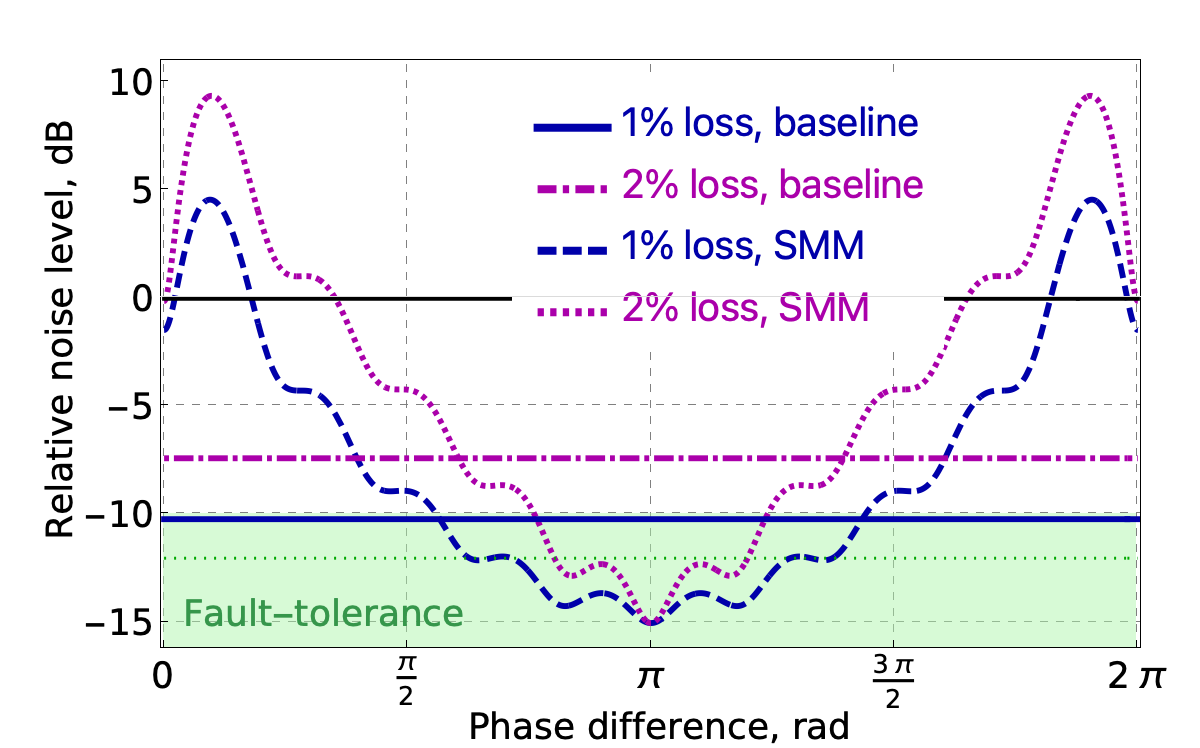}
  \end{minipage}
  \caption{Left: a selected path of squeezed field through the quantum-correlated network, experiencing SMM at each node. Right: The effect of hot SMM on a squeezed state in a quantum network compared to the baseline where losses are added incoherently. Noise reduction below the shot noise level is shown as a function of the common phase at each step, assuming equal coupling strength and phase shift at each component. Any squeeze value above the threshold of -10\,dB is not useful for computation, which limits the range of possible phases. Two regimes are shown: for 1\,\% mismatch (blue) the baseline incoherent loss model predicts sufficient squeezing for fault-tolerant CVQC (green area), while 2\,\% mismatch (magenta) does not allow fault-tolerance. For the hot SMM case, if the phases are not optimized, a significant part of the phase space results in insufficient squeezing for fault-tolerant CVQC, even where it would be allowed by the baseline loss model. At the same time, by optimizing the phases, fault-tolerant CVQC can be achieved even where it would not be expected within the incoherent model.}
  \label{fig:theory_network}
\end{figure}

We extend the simplified model from the previous section to analyze the effect of the mismatch on one path in the large-scale quantum network. We consider a series of mixing components, each defined by the coupling strength $k_i$ and the differential phase $\phi_i$.
While the realistic experimental setups may include different coupling strength for each component, and a variety of overall phase shifts, we can provide the envelope of the expected effect by assuming equal coupling strength $k_i = k$ and equal phase shifts $\phi_i = \phi$. The full derivation is provided in the Supplementary Material.

Fig.\,\ref{fig:theory_network} shows the expected squeezing level as a function of the common phase shift for two different mode mismatch values: 1\,\% and 2\,\% per node. For the first case, the incoherent account for the mismatches would result in 10.2\,dB of squeezing after 10 nodes, which is sufficient for fault-tolerant CVQC. However, the hot SMM analysis shows that for about 55\,\% of the phase values, the output squeezing drops below 10\,dB, making it unsuitable for fault-tolerant CVQC without careful phase design. At the same time, this phase optimization can allow to achieve fault tolerance where it would not be expected within the incoherent model, as illustrated by  the second case of 2\,\% mismatch per node. In this case, the incoherent model predicts 7.4\,dB of squeezing after 10 nodes, which is insufficient for fault-tolerant CVQC. However, by tuning the differential phases between the modes, the output squeezing can be stabilized below 10\,dB for 25\,\% of the phase values.
This serves as an intuitive illustration of the challenges that may arise in large-scale quantum networks, where multiple mode-mixing components are present. It also highlights the opportunities that arise when SMM is taken into account.  In practice, a full modeling of the network is required to accurately predict the effect of hot SMM on the quantum states.

\backmatter

% \bmhead{Supplementary information}

\bmhead{Acknowledgements}
This research has been funded by the German Federal Ministry of Research, Technology and Space (BMFTR) (JG), grant no. 05A20GU5 and by ERDF of the European Union and by ``Fonds of the Hamburg Ministry of Science, Research, Equalities and Districts (BWFGB)'' (SG). The work of RS and MK was supported by the Deutsche Forschungsgemeinschaft (DFG) under Germany's Excellence Strategy EXC 2121 ``Quantum Universe''-390833306.

%This work was supported by...
%During preparation of this work, we have become aware about the related recent work further extending the study of\,\cite{McCuller2021} in the context of gravitational-wave detection.

\section*{Declarations}
The authors declare no competing interests.
The data that support the findings of this study are available from the corresponding author upon reasonable request.

\textbf{Authors contributions according to the CRediT system:}

SG: \textbf{lead} on data curation, formal analysis (data); \textbf{equal contributions} on software, investigation (conducting the experiment), writing - review \& editing; \textbf{supporting contributions} on methodology and writing - original draft.

JG: \textbf{lead} on investigation (experimental design); \textbf{equal contributions} on investigation (conducting the experiment), visualization, writing - original draft and writing - review \& editing; \textbf{supporting contributions} on data curation, software, methodology and project administration.

RS: \textbf{lead} on funding acquisition and resources; \textbf{equal contributions} on supervision, writing - review \& editing; \textbf{supporting contributions} on conceptualization, methodology and project administration.

MK: \textbf{lead} on conceptualization, formal analysis (theory), methodology and project administration; \textbf{equal contributions} on software, supervision, visualization, writing - original draft and writing - review \& editing, \textbf{supporting contributions} on formal analysis (data).

% \begin{itemize}
% \item Conflict of interest/Competing interests (check journal-specific guidelines for which heading to use)
% \item Ethics approval and consent to participate
% \item Consent for publication
% \item Materials availability
% \end{itemize}

% \noindent
% If any of the sections are not relevant to your manuscript, please include the heading and write `Not applicable' for that section. 

% \bigskip

% Springer journals and proceedings: \url{https://www.springer.com/gp/editorial-policies}

% Nature Portfolio journals: \url{https://www.nature.com/nature-research/editorial-policies}

% \textit{Scientific Reports}: \url{https://www.nature.com/srep/journal-policies/editorial-policies}

% BMC journals: \url{https://www.biomedcentral.com/getpublished/editorial-policies}
% \end{flushleft}
% ?


\bibliography{sn-bibliography}% common bib file

\end{document}